\documentstyle[psfig,twoside,fleqn,espcrc2]{article}
\newcommand{\AmS}{{\protect\the\textfont2
  A\kern-.1667em\lower.5ex\hbox{M}\kern-.125emS}}

\title{Numerical Analysis of the Double Scaling Limit in the bosonic
part of the IIB Matrix Model
\thanks{presented by S.Horata}}

\author{S.Horata
        \address{Theory Division, Institute of Particle and Nuclear Studies,
        KEK, High Energy Accelerator Research Organization,
        Tsukuba, Ibaraki 305-0801, Japan
        }
        ,
        H.S.Egawa
		$^{{\rm a} ,}$
        \address{Department of Physics, Tokai University, 
        Hiratsuka, Kanagawa 259-1292, Japan}
		}
\begin{document}

\begin{abstract}
The bosonic IIB matrix model, which contains the bosonic part of the IIB
matrix model conjectured to be a non-perturbative definition of the
type IIB superstring theory, is studied using a numerical method.
The large $N$ scaling behavior of the model is shown performing a Monte
Carlo simulation.
The expectation value of the Wilson loop operator is measured and the
string tension is estimated.
From the numerical results, the prescription of the double scaling limit 
 is obtained.
\end{abstract}
\maketitle 
%%%%%%%%%%%%%%%%%%%%%%%%%%%%%%%%%%%%%%%%%%%%%%%%%%%%%%%%%%%%%%%%%%%%%%%%%%%%
\section{Introduction} 
%%%%%%%%%%%%%%%%%%%%%%%%%%%%%%%%%%%%%%%%%%%%%%%%%%%%%%%%%%%%%%%%%%%%%%%%%%%%
Recently, several models have been proposed as non-perturbative
formulations of string theory\cite{BFSS,IKKT,AIKKTT,KAWAI}.
The IIB matrix model\cite{IKKT}, which is zero-volume limit of the
ten-dimensional large $N$ supersymmetric Yang-Mills theory, has been
proposed as a constructive definition of the type IIB superstring theory.
It is defined by the following action,
\begin{equation}
 S = - \frac{1}{g^2} {\rm tr} \left( \frac{1}{4}[A_\mu, A_\nu]^2 
     + \frac{1}{2} \bar{\psi} \Gamma^\mu [A_\mu,\psi]\right) ,
\end{equation}
where $A_\mu$ and $\psi$ are $N \times N$ traceless Hermitian matrices. 
The interesting feature of this model is that the space-time coordinates
are considered as the eigenvalues of these matrices.
Then, we expect that the fundamental issues including the dimensionality
and the quantum gravity can be understood by studying the dynamics of
the model.

To take the continuum limit ($N \rightarrow \infty$) for the IIB Matrix 
model, a sensible double scaling limit should be determined dynamically
for the scaling property of the two important quantities of the model,
the string scale ($\alpha'$) and the string coupling constant
($g_{str}$)\cite{FKKT}.
\begin{eqnarray}
 \alpha' &\sim& g^2 N^{a+b} = g^2 N^{\gamma} \sim Constant, \nonumber \\
 g_{str} &\sim& N^{a-b},
\end{eqnarray}
where one have a restriction, $a=b$ for the finite value of the string
coupling constant ($g_{str}$).

For studying the dynamical aspects of the model, some Monte Carlo
simulations have been performed.
In Ref.\cite{HNT}, the bosonic model has been studied with a analytical
method as well as a numerical one.
The numerical results support the $1/D$ expansion as an effective tool
detecting the large $N$ scaling behavior of the model.
The leading term of the $1/D$ expansion at $D>3$ suggests that the
exponent ($\gamma$) takes a value of 1.

In analogy with the two-dimensional model, the Eguchi-Kawai model, 
we study the scaling property of the Wilson loop in ten-dimension. 
The area law of the Wilson loop operator has been found in the
four-dimensional model in Ref.\cite{AABHN}.
Then we consider that the scaling property of the string tension is
estimated in ten-dimensional model with investigating the scaling
property of the Wilson loop in ten-dimension. 

In this article, we present the data which show the existence of the
double scaling limit in the model and discuss our numerical results.

%%%%%%%%%%%%%%%%%%%%%%%%%%%%%%%%%%%%%%%%%%%%%%%%%%%%%%%%%%%%%%%%%%%%%%%%%%%%
\section{Monte Carlo simulation and scaling behavior}
%%%%%%%%%%%%%%%%%%%%%%%%%%%%%%%%%%%%%%%%%%%%%%%%%%%%%%%%%%%%%%%%%%%%%%%%%%%%
For numerical simulation of the model, we consider a bosonic version of
the IIB matrix model, namely a model which can be obtained from the IIB
matrix model by dropping the fermions by hand\cite{HNT}.
In the IIB matrix model, supersymmerty is expected to be very important
as is discussed in many places of the
literature\cite{KAWAI,AABHN,AABHN2,AABHN3}. 
However, it is difficult to calculate with numerical method. 
In ref.\cite{AABHN2}, numerical study of a supersymmetric version of
the IIB matrix model is discussed.
In this article, we study the most simplest model.
The bosonic model of the IIB matrix model is given by
\begin{equation}
 S_{bosonic} = -\frac{1}{g^2} {\rm tr} [A_\mu, A_\nu][A^\mu,A^\nu], \label{action}
\end{equation}
where $A_\mu$ are $N \times N$ Hermitian matrices representing the
ten-dimensional gauge fields.
The coupling constant ($g$) is nothing but a scale parameter and is
absorbed with the rescaling of the gauge field ($A_\mu$) as $A_\mu
\rightarrow \frac{1}{\sqrt{g}}A_\mu$.

we consider the partition function of bosonic model in
ten-dimension,
\begin{equation}
 Z = \int dA e^{-\frac{1}{g^2} {\rm tr} [A_\mu, A_\nu][A^\mu,A^\nu]},
\end{equation}
where the measure of gauge fields is defined by
\begin{equation}
 dA = \prod_{\mu=1}^{10} \left[ \prod_{i=1}^{N} \prod_{j=i}^{N}
 d(A_{\mu}^{ij}) \right].
\end{equation}

The action is quadratic with respect to each component, which means that
we can update each component by generating gaussian random number in the
heat-bath and the Metropolis algorithm.

To investigate the scaling behavior of the string scale ($\alpha'$), 
we study the Wilson loop operator in the IIB matrix model.

The Wilson loop operator and the large $N$ behavior have been studied 
with the light-cone string field theory of the type IIB
superstring\cite{FKKT}.
The Wilson loop operator ($w(C)$) is defined as,
\begin{equation}
 w(C) = {\rm tr}(v(C)),
\end{equation}
where $C$ denotes the closed path and $v(C)$ is defined as $v(C) = U_\mu
\cdots U_\mu =  P_C \exp (i \oint d \sigma k_\mu A_\mu) $ in the
bosonic model. 
The matrices ($U_\mu$) are considered as the unitary matrices,
\begin{equation}
 U_\mu = \exp ( i \int d l A_\mu).
\end{equation}

In the ordinary lattice gauge field theory, the expectation value of the
Wilson loop operator which spreads a large area behaves as follows,
\begin{equation}
 w(I,J) \sim \exp (- K I\times J),
\end{equation}
where $I$ and $J$ are the side lengths of the rectangular loop and $K$
denotes the string tension.
In the same analogy, we study the Wilson loop operator of the bosonic
model of the IIB matrix model.

From the scaling relation of two-dimensional Eguchi-Kawai model,
\begin{equation}
 g^2 N \sim Constant,
\end{equation}
It is expected that the similar scaling relation also holds in the IIB matrix
model as\cite{FKKT}
\begin{equation}
 \alpha' \sim g^2 N^{\gamma} \sim Constant. \label{IIB-SCALING}
\end{equation}
The exponent ($\gamma$) should be determined dynamically from
the model.
For the large $N$ limit, the parameter ($g^2 N^{\gamma}$) must be fixed
in the IIB matrix model in the same manners as the parameter ($g^2 N$) must
be fixed in the Eguchi-Kawai model.

It is considered that the bosonic model is equivalent to the $D>2$
Eguchi-Kawai model in the weak coupling limit.
Since the $U(1)^D$ symmetry rotates all the eigenvalues by the same
angle, the following expansion is valid in the weak coupling region,
\begin{equation}
 U_\mu \sim e^{i \alpha_\mu} e^{i A_\mu},
\end{equation}
where $\alpha_\mu$ take constant values due to the $U(1)^D$ symmetry and
$A_\mu$ are small.
The bosonic model action can be obtained by expanding the action of 
the ten-dimensional Eguchi-Kawai model in terms of $A_\mu$.
When the higer order terms of $A_\mu$ can be neglected, we can obtain
the area law of the Wilson loop operator as
\begin{eqnarray}
 w(I \times J) &=& <{\rm tr}(U_\mu \cdots U_\mu)_{I \times J}> \\ \nonumber
               &\simeq& <{\rm tr}(e^{i A_\mu} \cdots e^{i A_\mu})_{I \times J}> \\ \nonumber
               &\sim& \exp ( -K (I \times J)). \label{IIB-AREALAW}
\end{eqnarray}
In Ref.\cite{AABHN}, the area dependence of the Wilson loop operator has 
been measured and found the area law in four-dimension, $D=4$.

We calculate the eigenvalue of Wilson loop operator numerically in
ten-dimension, $D=10$,
\begin{equation}
 w(C) = <\frac{1}{N} {\rm tr} (e^{i A_\mu} \cdots e^{i A_\nu}) >.
\end{equation}
We take the loop ($C$) as the rectangular ($I \times J$) where we select 
any two direction ($\mu,\nu$) in ten-dimension. 
We consider that in the bosonic model the isotropy of the
ten-dimensional space-time is not broken down spontaneously,
since the dependence of the direction of the rectangular is not found.

From the numerical results, the Wilson loop operator closes to
exponential curve with the large size area as
\begin{equation}
 w(I,J) \sim \exp (- K I\times J),
\end{equation}
where $K$ denotes string tension.
We plot the results in Fig.1.
%
%%%%%%%%%% the string tension %%%%%%%%%%%%%%
\begin{figure}
\centerline{\psfig{file=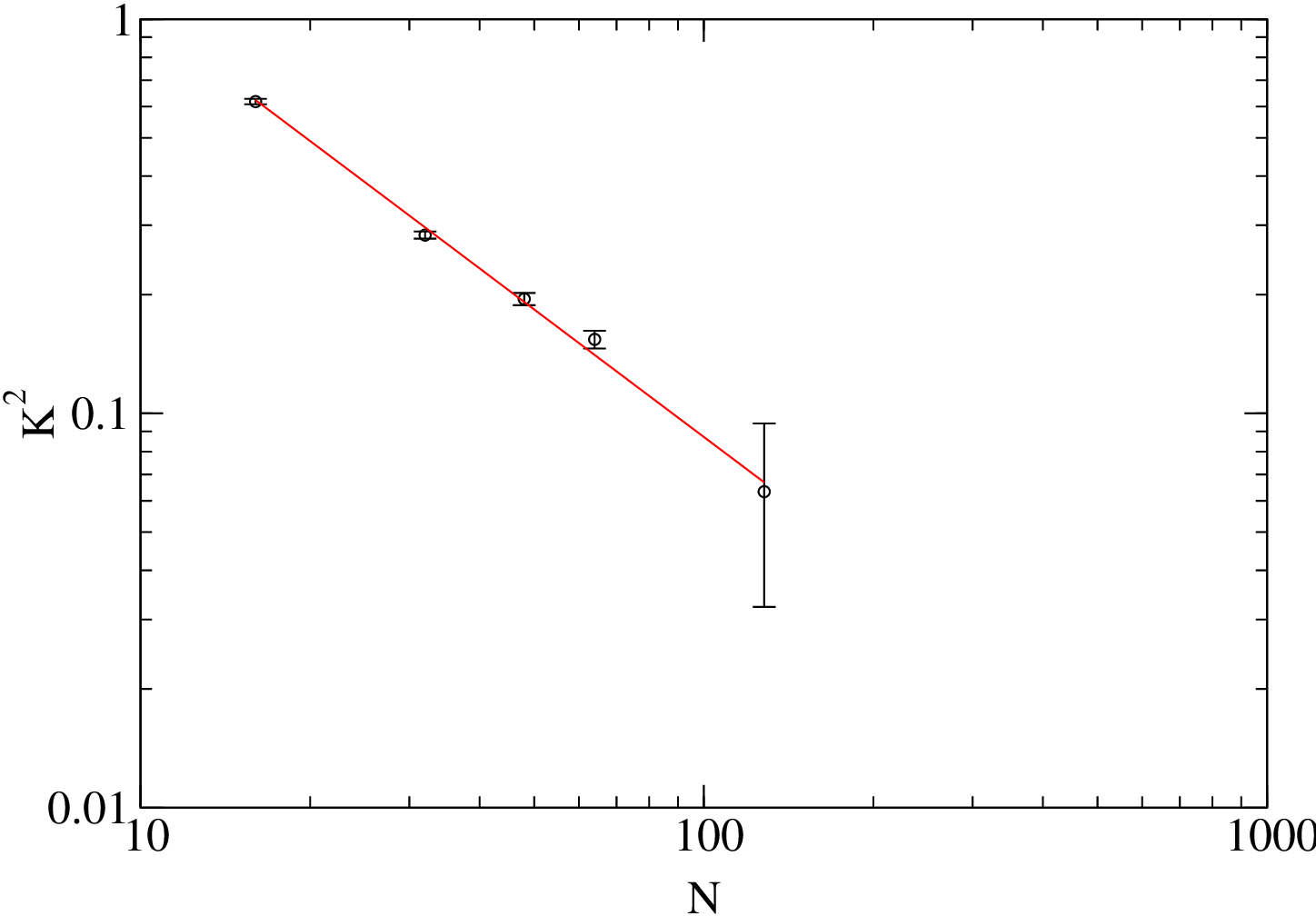,height=6cm,width=8cm}}
\label{fig:kappa}
\begin{center}
{\small
Fig.1 The measurement results of the string tension $a$.
}
\end{center}
\end{figure}
%%%%%%%%%%%%%%%%%%%%%%%%%%%%%%%%%%%%%%%%%%%%%%%%%%%%%%%
%
Then we can estimate the string scale
($\alpha'$) as 
\begin{equation}
 \alpha'^2 \sim 1/K^2 \sim g^{2} N^{1.07(1)} = Constant. \nonumber
\end{equation}
%
%%%%%%%%%%%%%%%%%%%%%%%%%%%%%%%%%%%%%%%%%%%%%%%%%%%%%%%%%%%%%%%%%%%%%%%%%%%%
\section{Summary}
%%%%%%%%%%%%%%%%%%%%%%%%%%%%%%%%%%%%%%%%%%%%%%%%%%%%%%%%%%%%%%%%%%%%%%%%%%%%
%
Let us summarize the main points made in our calculation.
We confirm that the Wilson loop operator in the ten-dimensional bosonic
model obeys the area law.
For the scaling behavior of the bosonic model of the IIB matrix model,
our numerical estimation is
\begin{equation}
 \alpha'^2 \sim g^{2} N^{1.07(1)} = Constant.
\end{equation}
Furthermore the numerical result suggests that the large $N$ behavior of
the square root of the string tension approximately equal to the inverse of the
extent of the space-time\cite{HNT,HE}.
\begin{equation}
 K^{1/2} \sim g^{-1} N^{-0.54(1)} \sim R^{-1}.
\end{equation}
It means that the Planck scale of the theory has the same scaling
property of the extent of the space time in ten-dimension.
The numerical result is consistent to the suggestion from 
the string field theory on light-cone frame\cite{FKKT} and the $1/D$
expansion\cite{HNT}.


\begin{thebibliography}{99}
%%%%%%%%%%%%%%%%%%%%%%%%%%%%%%%%%%%%%%%%%%%%%%%%%%%%%%%%%%%%%%%%%%%%%%%%%%%%
%%%%%%%%%%%%%%%%%%%%%%%%%%%% IIB Matrix Model %%%%%%%%%%%%%%%%%%%%%%%%%%%%%%
\bibitem{BFSS} T.Banks, W.Fischler, S.H. Shenker and L.Susskind,
		Phys. Rev. {\bf D55} (1997) 5112, hep-th/9612115. 
\bibitem{IKKT} N.Ishibashi, H.Kawai, Y.Kitazawa and A.Tsuchiya,
		Nucl. Phys. {\bf B498} (1997) 467.
\bibitem{AIKKTT} H. Aoki, S. Iso, H. Kawai, Y. Kitazawa, T. Tada,
		A. Tsuchiya, Prog.Theor.Phys.Suppl. 134 (1999) 47-83.
\bibitem{KAWAI} H.Kawai, talked at Lattice2000.
\bibitem{FKKT} M.Fukuma, H.Kawai, Y.Kitazawa and A.Tsuchiya,
		Nucl. Phys. {\bf B510} (1998) 158.
\bibitem{HNT} T.Hotta, J.Nishimura and A.Tsuchiya, Nucl. Phys. {\bf
		B545} (1995), 543.
\bibitem{AABHN} J.Ambj{\o}rn, K.N.Anagnostopoulos, W.Bietenholz,
		T.Hotta and J.Nishimura, hep-th/0003208.
\bibitem{AABHN2} J.Ambj{\o}rn, K.N.Anagnostopoulos, W.Bietenholz,
		T.Hotta and J.Nishimura, hep-th/0005147.
\bibitem{AABHN3} J.Ambj{\o}rn, K.N.Anagnostopoulos, W.Bietenholz,
		T.Hotta and J.Nishimura, hep-lat/0009030 (talked at Lattice2000)
\bibitem{HE} S.Horata and H.S.Egawa, hep-th/0005157. 

\end{thebibliography}
\end{document}